\documentclass[epjCONF,columns]{svjour}
\usepackage{graphics}
\usepackage[varg]{txfonts} % Times fonts
\usepackage[latin1]{inputenc}
\session-title{Conference Title, to be filled}
\begin{document}
\title{Prompt Emission from Tidal Disruptions of White Dwarfs by Intermediate Mass Black Holes}

\author{Roman V. Shcherbakov\inst{1,2}\fnmsep\thanks{\email{roman@astro.umd.edu}} \and Asaf Pe'er\inst{3} \and  Christopher S. Reynolds\inst{1}
\and Roland Haas\inst{4,5} \and Tanja Bode\inst{5} \and Pablo Laguna\inst{5}}
\institute{Department of Astronomy, University of Maryland, College Park, MD 20742-2421, USA \and
Hubble Fellow \and
Harvard-Smithsonian Center for Astrophysics, 60 Garden Street, Cambridge, MA 02138, USA \and
Theoretical AstroPhysics Including Relativity, California Institute of Technology, Pasadena, CA 91125, USA \and
Center for Relativistic Astrophysics, School of Physics, Georgia Institute of Technology, Atlanta, GA 30332, USA}
\abstract{We present a qualitative picture of prompt emission from tidal disruptions of white dwarfs (WD)
by intermediate mass black holes (IMBH). The smaller size of an IMBH compared to a supermassive black hole and a smaller tidal radius of a WD disruption lead to a very fast event with
high peak luminosity. Magnetic field is generated in situ following the tidal disruption, which leads to effective accretion.
Since large-scale magnetic field is also produced, geometrically thick super-Eddington inflow leads to a relativistic jet.
The dense jet possesses a photosphere, which emits quasi-thermal radiation in soft X-rays. The source can be classified as a long low-luminosity gamma-ray burst (ll-GRB).
Tidal compression of a WD causes nuclear ignition, which is observable as an accompanying supernova.
We suggest that GRB060218 and SN2006aj is such a pair of ll-GRB and supernova.
We argue that in a flux-limited sample the disruptions of WDs by IMBHs are more frequent then the disruptions of other stars by IMBHs.}
\maketitle
\section{Introduction}
\label{intro}
Two well-known populations of BHs are the stellar mass BHs with mass $M_{BH}<100M_{\rm Sun}$ and supermassive black holes (SMBH) with mass $M_{BH}>10^5M_{\rm Sun}$.
A third population of intermediate mass black holes (IMBH) likely exists with masses $100M_{\rm Sun}<M_{BH}<10^5M_{\rm Sun}$.
They could live in the centers of galaxies that fail to feed their BHs \cite{Dong:2007dg} or in globular clusters \cite{Fabbiano:1997yt,Fabbiano:2001fr,Colbert:1999fe,Matsumoto:2001oi,Gultekin:2004gh}. IMBH can form as a result of a collapse of a massive star \cite{Fryer:2001fe,Madau:2001po,Schneider:2002dp} or a massive cloud \cite{Begelman:2006jk} or grow from a stellar mass BH. While the presence of stellar mass BHs and SMBHs
is established, only tentative candidates of IMBHs exist \cite{Gebhardt:2002hj,Dong:2007dg,Davis:2011ka} and the debates of the nature of the candidates are ongoing (e.g. \cite{Baumgardt:2003wd}). More IMBH candidates with qualitatively different observational signatures could help identify those objects.

Tidal disruptions of stars by IMBHs could provide such a qualitatively different signature.
Soon after the disruption, the isotropic luminosity as well as the intrinsic luminosity from the system may greatly exceed the Eddington limit owing to jet emission.
A jet naturally appears in thick magnetized accretion disks, especially those with strong mean poloidal field \cite{McKinney:2009rt,McKinney2012}.
In this work we briefly summarize how the jet can be launched following a tidal disruption of a WD by an IMBH.
We qualitatively describe the radiative signatures of a jet and the temporal behavior.
Tidal compression of a WD may lead to thermonuclear ignition.
Thus the source may appear as a weak supernova simultaneous with a low-luminosity gamma-ray burst (ll-GRB).
The tidal disruptions of MS stars by IMBHs happen more frequently, than disruption of WDs by IMBHs.
Nevertheless, significantly larger maximum radiation power from WD-IMBH disruption could make those sources easier to find.
We identify GRB060218 and an associated supernova SN2006aj as a WD-IMBH tidal disruption candidate.

\begin{figure*}[htbp]
    \centering\includegraphics{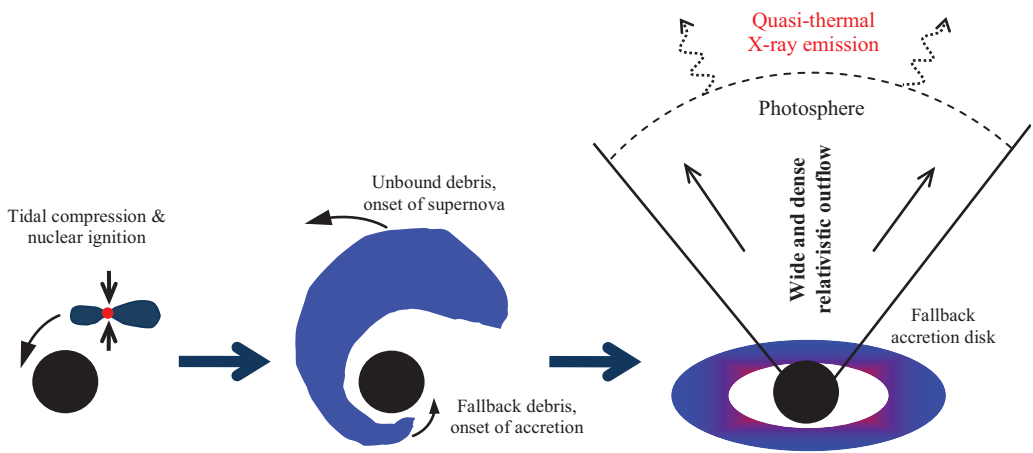}
    \caption{Production of prompt emission in tidal disruptions of WDs by IMBHs. Extreme tidal compression in the plane perpendicular to the disruption plane leads to nuclear ignition.
    The outflowing debris produce a supernova signature. A fraction of debris may still remain bound to the BH and form an accretion disk.
    Thick fallback accretion disk generates the magnetic field and creates super-Eddington inflow. Whenever the large-scale magnetic field is present, a part
    of the inflow get accelerated in a relativistic jet. The dense outflow is optically thick to Compton scattering. Quasi-thermal emission is produced from the jet photosphere.}
    \label{fig:scheme}
\end{figure*}

\section{Jets in tidal disruptions}
\label{sec:1}
Jet, a relativistic outflow, is a ubiquitous feature of astrophysical accretion flows.
Jets are observed in X-ray binaries \cite{Remillard:2006} and active galactic nuclei \cite{Boettcher:2012}.
Fast relativistic outflow are inferred in GRBs \cite{Meszaros:2002re,Meszaros2006,Piran2004,Zhang:2007re}.
A tidal disruption event Swift J1644+57 is thought to have a jet \cite{Bloom11,Burrows11,Levan2011,Zauderer11}.
A fundamental question about jets is how to launch them.

Two main mechanisms to drive relativistic outflows are magnetic launching and fireball launching.
A large-scale poloidal magnetic field is thought to be the main ingredient in magnetic launching scenario \cite{McKinney:2009rt,McKinney2012}.
Such magnetic field could be brought in by large magnetized clouds.
Alternatively, amplification of magnetic field may produce strong poloidal component via dynamo action \cite{Brandenburg:2005pr}.
The small-scale field is generated on the dynamical timescale via magneto-rotational instability (MRI) \cite{Balbus:1991jk,Balbus:1998fe}.
Exponential amplification by a factor of $10^6$ requires $\sim50$ orbital periods $t_{\rm dyn}$ in a Keplerian MRI-unstable turbulent flow \cite{Stone:1996ed}.
The large-scale field generation via the dynamo action is a process, which operates on a slower resistive timescale
\cite{Brandenburg:2005pr}. Nevertheless, thick magnetized disks could have the viscous timescale on the order of $100t_{\rm dyn}$.
Then $200t_{\rm dyn}$ is all it takes to produce a strong large-scale magnetic field.
The rotation in the system with large-scale poloidal field creates an outflow. The rotation of the BH leads to jets via Blandford-Znajek process \cite{Blandford1977},
while the accretion disk rotation induces Blandford-Payne mechanism \cite{Blandford1982}.

The fireball model \cite{Goodman:1986,Paczynski:1986,Paczynski1994} was devised in applications to GRBs.
The extreme neutrino flux produced in a collapse of a massive stellar core can accelerate a small amount of material to relativistic velocities.
However, the neutrino flux from WD-IMBH tidal disruption is small, since temperature and particle density reach only the values of $10^{9.5}$~K and $10^{29}{\rm cm}^{-3}$,
respectively \cite{Haas:2012ak}. Thus, fireball jet launching is not an option for tidal disruptions of WDs.

The magnetic field in the fallback debris after WD-IMBH tidal disruption is given approximately by the magnetic field of a WD, since its radius $r_{WD}$
is about the IMBH Schwarzschild radius $r_s$. The typical WD magnetic field is $B\lesssim10^4$~G \cite{Putney:1999dg},
while some objects have a field up to $B\sim10^8$~G \cite{Angel:1978yt}. The equipartition magnetic field is $B\sim10^{10-11}$~G
for accretion rate $\dot{M}\sim10^{4}M_{\odot}{\rm yr}^{-1}$ predicted for WD-IMBH events \cite{Haas:2012ak}.
Even though the equipartition magnetic field is much larger, it may take only $\sim200t_{\rm dyn}$ after the fallback disk starts forming to generate strong large-scale magnetic field and launch the jet. As noted by \cite{Bloom11}, the initial magnetic field following the disruption in Swift J1644+57 could have been substantially weaker than the equipartition value.
Then the similar process could have been responsible for magnetic field amplification and production of a strong jet in that source.

The jet Lorentz factor has a crucial influence on the emitted radiation. GRB jets have high Lorentz factors of up to $\Gamma\sim 1000$ (see \cite{Lu:2012de} for the review),
while AGN jets in are estimated to move at $\Gamma\sim10$ (see e.g. \cite{Henri:2006hj}). Different amounts of external pressure support might be responsible for the difference.
GRB jets following the collapse of massive stars are found to effectively accelerate matter in a magnetically-dominated outflow with substantial external pressure support from a surrounding star \cite{Tchekhovskoy2010,Tchekhovskoy2011}. The pressure support is much weak in other sources, and the Lorentz factor stays low.
Since there is no material surrounding the tidal disruption region prior to the event, no substantial pressure support is expected for the resulting jet.
Then tidal disruption jet should have a Lorentz factor of $\Gamma\sim10$. The bulk Lorentz factor of $\Gamma=10-20$ estimated from observations of Swift J1644+57 \cite{Metzger11}
is consistent with AGN-like jet.

In sum, the material returns back to the IMBH at a super-Eddington fallback rate.
After an initial period of strong magnetic field generation, the magnetic stresses ensure that matter loses the angular momentum
and accretes onto the black hole. The accretion rate then follows the fallback rate.
The jet is launched whenever the strong poloidal large-scale magnetic field is generated.
The presence of a relativistic outflow dramatically alters the radiative signature of the disruption.

\section{Radiation from the jet}
\label{sec:2}
A variety of jet emission models exist for different types of jets.
They can be generally divided into two broad categories based on particle density.
At low density the jet is optically thin to Compton scattering and absorption with total optical depth $\tau<1$, while seed photon production in a jet is relatively inefficient.
External Comptonization of the accretion disk photons and the reprocessed photons plays a major role.
AGN jets including blazar jets are found to operate in a low-density regime \cite{Ghisellini2009}.

At high density the jet is optically thick to Compton scattering and absorption with $\tau>1$.
The production of seed photons is more efficient, and the external photons cannot penetrate through the jet.
Such high-density jet would have a larger high-energy contribution from Comptonization of its own photons, e.g., via synchrotron-self-Compton (SSC) mechanism.
An important other feature would be the presence of a photosphere at a distance $r_{\rm ph}$ from the BH along the jet.
The photons at $r<r_{\rm ph}$ can effectively interact with the electrons and thermalize.
Thermal black body emission is expected from the photosphere. GRBs are found to operate in a high-density regime \cite{Meszaros2006}.
Thermal photospheric emission was observed in several GRBs \cite{Peer:2007sd,Fan:2012th}.

The tidal disruption jet would naturally have a high particle density, especially after the disruption of a WD by an IMBH.
The low-density model based on external Comptonization might not be immediately applicable to these objects.
Instead, tidal disruption jets should resemble GRB jets.
However, a larger BH mass and a smaller Lorentz factor modify the prompt emission of WD-IMBH disruption from that of a typical GRB jet.
The qualitative features of prompt emission are:
\begin{enumerate}
\item{the event duration is given by a large fallback time $t_{\rm fall}=10^3-10^4$~s,}
\item{relatively low accretion rate leads to low jet power and, correspondingly, low radiation power,}
\item{jet power may be lower than estimated from magnetic field equipartition, since the time to generate the large-scale magnetic field can be comparable to $t_{\rm fall}$,}
\item{low jet power leads to lower peak emission frequency, since it is determined by a magnetic field in a jet,}
\item{low Lorentz factor leads to softer emission.}
\end{enumerate}
In sum, WD-IMBH tidal disruption produces soft ll-GRB. The quantitative description of prompt emission is considered in the main paper (Shcherbakov et al., 2012, in prep.).

\section{Associated supernova explosion}
Early theoretical work found strong compression of a WD before the disruption along the orbital angular momentum axis \cite{Luminet:1985re,Luminet:1989th}.%Luminet:1985yt,
Such compression can lead to extreme temperature and density causing thermonuclear ignition.
Later numerical simulations with simplified nuclear network \cite{Rosswog:2008ie} confirmed the ignition.
They showed that vastly different explosion energies and final nuclear compositions are possible.
The explosions of heavy WDs coming close to the BH may appear more similar to Type Ia supernova.
Less massive WDs may have lower explosion energy and produce much less iron.
Those could be classified as underluminous fast Type Ib/c supernova, since no hydrogen is expected to remain on a WD and the silicon absorption line may be absent as well.
The nuclear energy release may be smaller than the release of gravitational energy in a disruption.
In such case the trajectories of debris are not strongly affected \cite{Haas:2012ak} and the maximum expansion velocity can reach $50,000{\rm km~s}^{-1}$.
In such WD supernova the ejecta mass should be less than Chandrasekhar mass.
However, the ejecta mass estimate depends on the inferred heating mechanism of supernova debris.

\section{Candidate sources}
\subsection{GRB 060218}
Since its launch in 2004, the \textit{Swift} satellite became an excellent tool to observe GRBs.
Several GRBs studied by the satellite have unusually long and soft emission. GRB 060218 at a redshift $z=0.033$ has the highest peak count rate of the sources with such features.
Despite triggering the BAT detector, most of the energy was emitted in soft X-rays \cite{Campana2006,Soderberg:2006na} over the event duration $t_{\rm dur}\sim2000$~s.
The source is modeled to have a thermal component with temperature about $T=0.2$~keV, less than a typical temperature of a thermal GRB component \cite{Campana2006,Butler:2007uy}. The source peak luminosity is $L\sim10^{47}{\rm erg~s}^{-1}$, which makes it very underluminous.
The \textit{Swift} XRT lightcurve is consistent with no fast variability.
The source exhibited fast flux decay after $2600$~s followed by the afterglow.
GRB 060218 has an associated fast underluminous supernova SN 2006aj, which was classified as Type Ib/c \cite{Soderberg:2006na,Maeda:2007gh,Mazzali:2007as}.
All features of the source can be qualitatively explained by a WD-IMBH tidal disruption.

Long event duration is consistent with a tidal disruption of a WD by a BH with mass $\sim10^4M_{\rm Sun}$, when the pericenter radius is about the tidal radius.
The observed X-rays can be interpreted as the blackbody radiation Comptonized by a thermal distribution of electrons.
The efficiency of Comptonization goes down with time. The observed thermal flux and the blackbody temperature imply a Lorentz factor of about several and jet base radius $10^{11}$~cm \cite{Peer:2007sd}, which leads to an independent BH mass estimate to also be $\sim10^4M_{\rm Sun}$.
The low luminosity of the source results from a jet with relatively large mass loading or substantially sub-equipartition large-scale magnetic field.
The smooth lightcurve is fully consistent with a relatively large size of the central engine.
The fast decay cannot be easily explained by the drop of the source flux. Instead, self-obscuration by cooled jet "exhaust" material at late times could be
responsible for steep flux decay. The period of steep flux decay smoothly connects to an afterglow, which is produced via the external shock.
The external shock is not absorbed by the cooling jet material. The afterglow may be partially powered by the central engine \cite{Fan:2006ag}.

The associated supernova was inferred to have a relatively low energy and ejecta mass $M_{\rm ej}\approx2M_{\rm Sun}$ \cite{Mazzali:2006na}.
The inferred ejecta mass is larger than the possible WD mass, which implies that either the exploding object is a core of a massive star or that $M_{\rm ej}$ was overestimated.
We argue that the latter could be the case. Supernova emission is typically powered by the decay of radioactive nickel.
The debris have low optical depth to radioactive decay products at the timescale of several days, when flux peaks.
Large ejecta mass is needed to intercept the decay products and convert the energy into the optical radiation.
The supply of energy from the central source at this late stage can lead to similar optical emission from ejecta with lower mass.
Substantial central source activity is expected in tidal disruptions at late times.
The evolution of the accretion disk may lead to shallow accretion rate temporal slope $\dot{M}\propto t^{-4/3}$ starting at
several hours after the disruption \cite{Cannizzo2009}. Then the accreted amount per logarithmic time interval behaves as $t^{-1/3}$, which is practically independent of time.

Despite the large amount of observational data for GRB060218 and the associated supernova, at least two more explanations were proposed for this source:
supernova shock breakout (e.g. \cite{Campana2006,Waxman:2007ap}) and jet from a newborn neutron magnetar \cite{Toma2007,Fan:2011jk}.  We describe those models and compare to our model in the main paper.

\subsection{Can disruptions of main sequence stars by IMBHs be observed?}
Apart from WDs, the IMBHs can readily disrupt the main sequence (MS) stars.
N-body modeling in globular clusters have shown that up to $15\%$ of total disruptions are those of WDs \cite{Baumgardt2004}, while the MS stars constitute the majority of disruptions.
Nevertheless, the peak accretion rate is much less for disruptions of MS stars. This leads to much lower peak luminosity and softer spectrum, assuming that jet launching is similar in WD and MS star disruptions. Then WD-IMBH encounters dominate the flux-limited sample of tidal disruptions by IMBHs.

\section{Acknowledgements}
The work is partially supported by NASA Hubble Fellowship grant HST-HF-51298.01.
\bibliographystyle{epj}

\end{document}